# Electric-Field Induced Phase Transitions in Capillary Electrophoretic Systems

Hakan Kaygusuz [1,2], F. Bedia Erim [3] and A. Nihat Berker [4,5]

[1] *Department of Basic Sciences, Faculty of Engineering and Natural Sciences, Altınbaş University, Mahmutbey, Istanbul 34218, Turkey*

[2] *Sabancı University SUNUM Nanotechnology Research Center, Istanbul 34956, Turkey*

[3] *Department of Chemistry, Faculty of Science and Letters, Istanbul Technical University, Maslak, Istanbul 34469, Turkey*

[4] *Faculty of Engineering and Natural Sciences, Kadir Has University, Cibali, Istanbul 34083, Turkey*

[5] *Department of Physics, Massachusetts Institute of Technology, Cambridge, Massachusetts 02139, USA*

**Abstract**

The movement of the particles in a capillary electrophoretic system under electroosmotic flow was modeled using Monte Carlo simulation with Metropolis algorithm. Two different cases, with repulsive and attractive interactions between molecules were taken into consideration. The simulation was done using a spin-like system where the interactions between the nearest and second closest neighbors were considered in two separate steps of the modeling study. A total of 20 different cases with different rate of interactions for both repulsive and attractive interactions were modeled. The movement of the particles through the capillary is defined as current. At a low interaction level between molecules, a regular electroosmotic flow is obtained, on the other hand, with increasing interactions between molecules the current shows a phase transition behavior. The results also show that a modular electroosmotic flow can be obtained for separations by tuning the ratio between molecular interactions and electric field strength.

## Introduction

Capillary electrophoresis (CE) is an electrophoretic separation technique based on the principle of electrophoretic mobility differences of analytes under an electric field. CE techniques include Capillary zone electrophoresis (CZE), capillary gel electrophoresis, isoelectric focusing, micellar electrokinetic chromatography, affinity capillary elecrophoresis, capillary electrochromatography are different modes of CE technique. CZE is the simplest and most common type of CE. CZE is known for its very low sample consumption, fast and efficient separation, and easy method development. It is used for a wide variety of sample types [1–10]. The separation occurs in fused silica capillary columns with micrometer-sized diameters, and a high voltage (in kilovolts) is applied between the electrodes. Charged analytes migrate under the electric field, and they separate according to their charges and sizes.

Monte Carlo simulations of capillary electrophoresis were reported for various purposes. The simulation for determining the entanglement-related bulk motion, probability, and duration of entanglement for a single flexible polyelectrolyte chain was reported previously [11]. Laminar and electroosmotic convection in capillary electrophoresis was studied [12] and it was found that the radial velocity profile has an important effect for nonretained solutes. Other studies include the simulation of the enantioseparation process [13] and reactive separations [14]. Many of these studies refer to the applications of capillary electrophoresis with different approaches by means of simulation lattices. Other mathematical modeling studies for the electrophoretic systems, such as the insulator based dielectrophoretic devices was recently reviewed by Hill and Lapizco-Encinas [15] or electromigration of electrolytes considering the deviation from electroneutrality [16].

The spin-1 Ising model $-\beta \mathcal{H} = \sum_{<ij>} J s_i s_j$ where i and j show the sites of the spin $s = \pm 1$ is very useful for its applicability in different physical systems with both ordering ($s_i$ = ±1) and non-ordering ($s_i$ = 0) degrees of freedom [17–19]. Spin-1 model has a single critical point on the temperature axis ($J^{-1}$). This model can be useful in modeling capillary electrophoretic systems as shown in this study. The main aim of this study is to model the movement of the particles in a capillary tube under electroosmotic flow and to find possible phase transitions in the current versus electric field strength in capillary electrophoresis with varying molecular interaction levels, based on a modified Ising model capillary column with Monte Carlo simulation.

**Theory**

The Monte Carlo model used in this study is written in the C & C++ language. A column with $N$ ($N ≥ 100$) lattice units of length and 10 lattice units of diameter was designed for the simulation. The first four layers of this column were initially filled with 100 randomly distributed, identical particles. Figure 1 shows an example to the first four layers of the simulation setup with six lattice units of diameter.

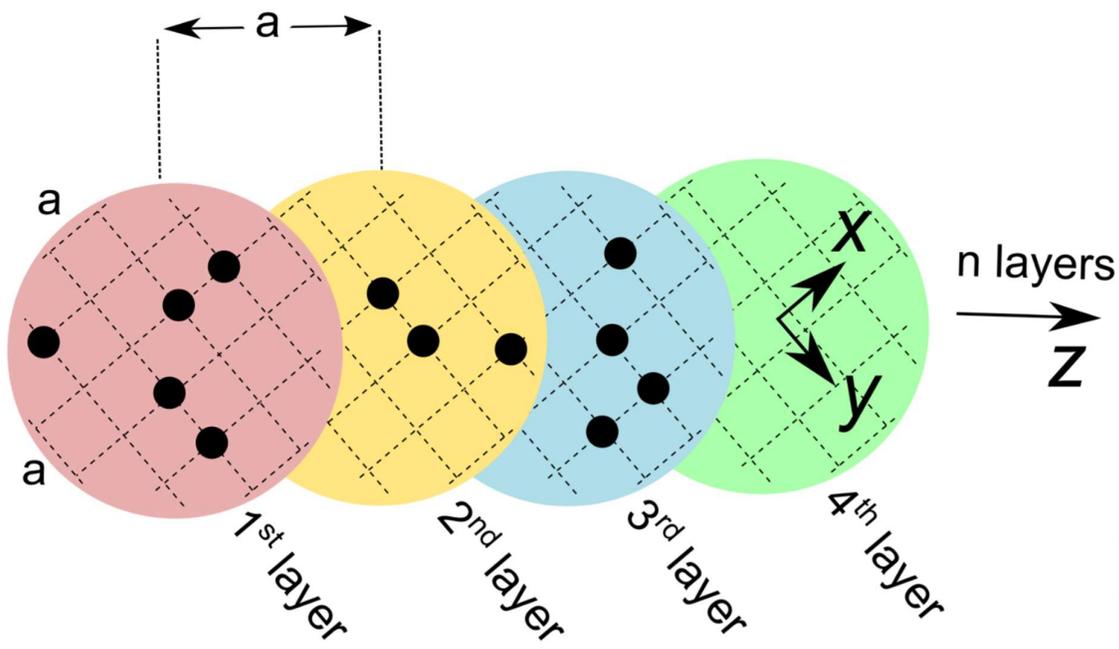

**Figure 1**. Schematic representation of a tube with 6 lattice units (a) of diameter. Each layer is separated by a. Black dots indicate randomly distributed particles.

The Monte Carlo (MC) method with the Metropolis algorithm was used in the simulation. 100 reorientation attempts are defined as 1 Monte Carlo step (MCS), which is used as the time unit. The simulation was performed for at least 300 MCS.

To calculate the effect of the increase in interaction between the particles, simulation was repeated for the following dimensionless $V$ values: 1, 3, 5, 7, 10, 25, 50, 100, and 500. In addition to these, one more simulation was performed by canceling the Metropolis criterion and this case is equivalent to $V = \infty$. Each simulation was done for repulsive and attractive interaction cases separately. The simulations were repeated at least 1000 times, and the average values were used in calculations.

The algorithm is as follows: At a certain value of electric field strength (E = 0.5 to 5), one particle is randomly selected from the lattice, and a random direction is assigned to this particle. There are 6 possible directions: Back and forth throughout the column ($\pm z$) and along column section ($\pm x$ and $\pm y$). If the direction leads to a forbidden movement, i.e., the particle is on the boundary of the column or the directed lattice point is already occupied by another particle, then this selection is omitted, and a new particle is randomly selected. If the direction leads to a possible movement, then this particle is moved for one unit to its new lattice position. The difference of the energy of the system ($\Delta G$) is related to the initial ($U_1$) and new ($U_2$) energies of the particle:

$$\Delta G = V(\Sigma U_2 - \Sigma U_1) + EV \qquad (1)$$

The second term ($EV$) is also considered as $\mu$:

$$\mu = qE \frac{a}{kT} \qquad (2)$$

Where $q$ is the charge of the particle, $E$ is the electric-field strength, and $a$ is the lattice constant, reflects the change in energy due to the work done by the electric field when the particle moves along or against the field in the $z$ direction. It is divided by $kT$, since all energies are scaled by temperature and therefore dimensionless in the Metropolis method.

The molecules are charged q and move under the influence of the electric field. The particle-particle interactions are the usual, short-range, water-shielded electric, spontaneous-induced dipolar, and quantum mechanical Pauli exclusion interactions.

The interactions between particles ($U$) were set +1 or -1 for attractive and repulsive interactions, respectively. For instance, if a particle has three neighbors in the attractive interaction, $\Sigma U$ = 3. The value of $E$ is positive +z direction and negative for -z direction. In the simulation if randomly selected particle stays in the same z-position and moves along the cross section, E is kept zero. The move is accepted when $\Delta G$ ≥ 0. Otherwise, exp($\Delta G$) needs to be compared to a random number $r$ (Metropolis algorithm). In this case, the probability of the acceptance p is:

$$p: \begin{cases} 1, e^{\Delta G} \geq r \\ 0, e^{\Delta G} < r \end{cases} \qquad (3)$$

Whether the move is accepted or not, the configuration is saved. Then another particle is selected and the whole process is repeated for 100 times to reach 1 MCS. After the simulation finishes for given $E$, the configuration is reset to the beginning for the next $E$ value.

The movement along z-axis is represented in Figure 2.

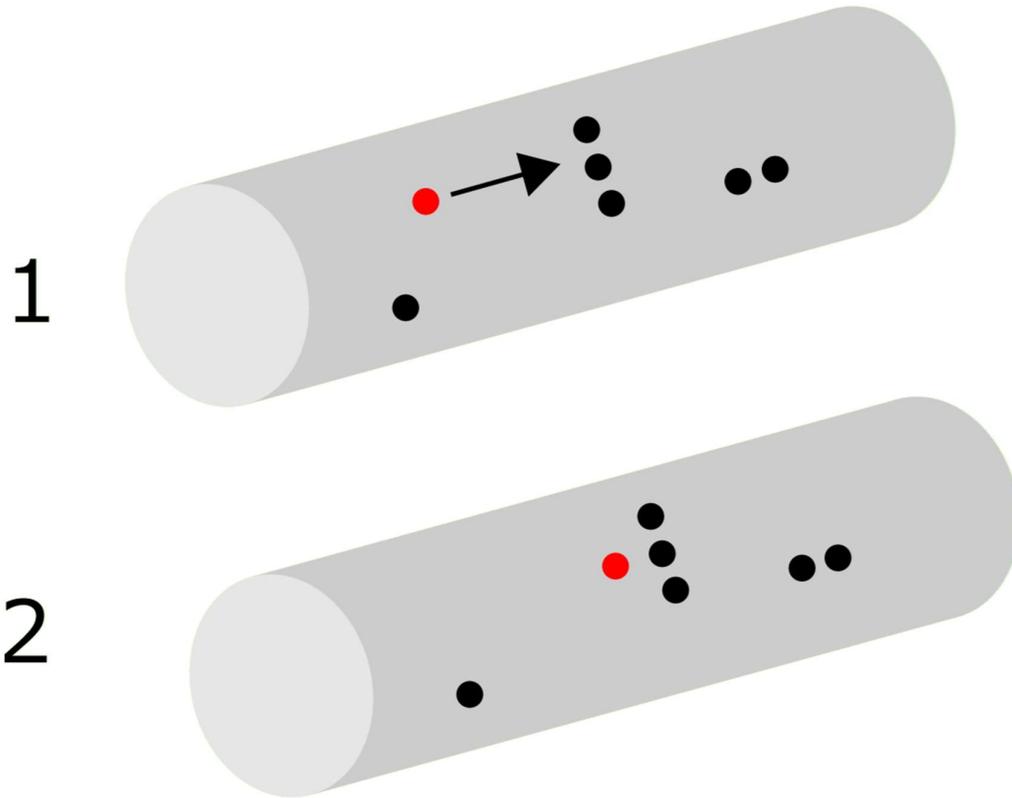

**Figure 2**. Schematic representation of a particle moving along z-axis. The increased amount of *E* forces the indicated particle to move forward in the repulsive case.

**Results and discussion**

Figure 3 shows the results of the simulations. The average number of particles traveled along the z axis per MCS is defined as the current (*J*), which can also be denoted as flow or migration.

Results indicate that at *V* = ∞, the current *J* shows a phase transition behavior with increasing electric field strength. The step functions are the standard signature of a phase transition [20]. This behavior is due to the interactions between particles. Particles with *n* neighbors require an electric field strength of *nE*. in order to break off from the neighbors and start the motion. Due to the random distribution of the particles used in this study, the largest jump is observed at *E* = 2. Another important result is the increase in the current when a repulsive interaction between the particles was set. As expected, particles tend to stay away from each other in this case. In other words, they repel each other and a higher *J* was obtained. A higher value of electric field strength is required to remove one of the clustered particles and move them along the tube for the attractive interaction case. On the other hand, in repulsive interaction, it is required to enforce the repelling particles to stay together and form a cluster. At $V = 1$, smooth curves were obtained, and the difference between the nature of the interaction becomes nonsignificant. At $V = 500$ the curve overlaps with $V = \infty$.

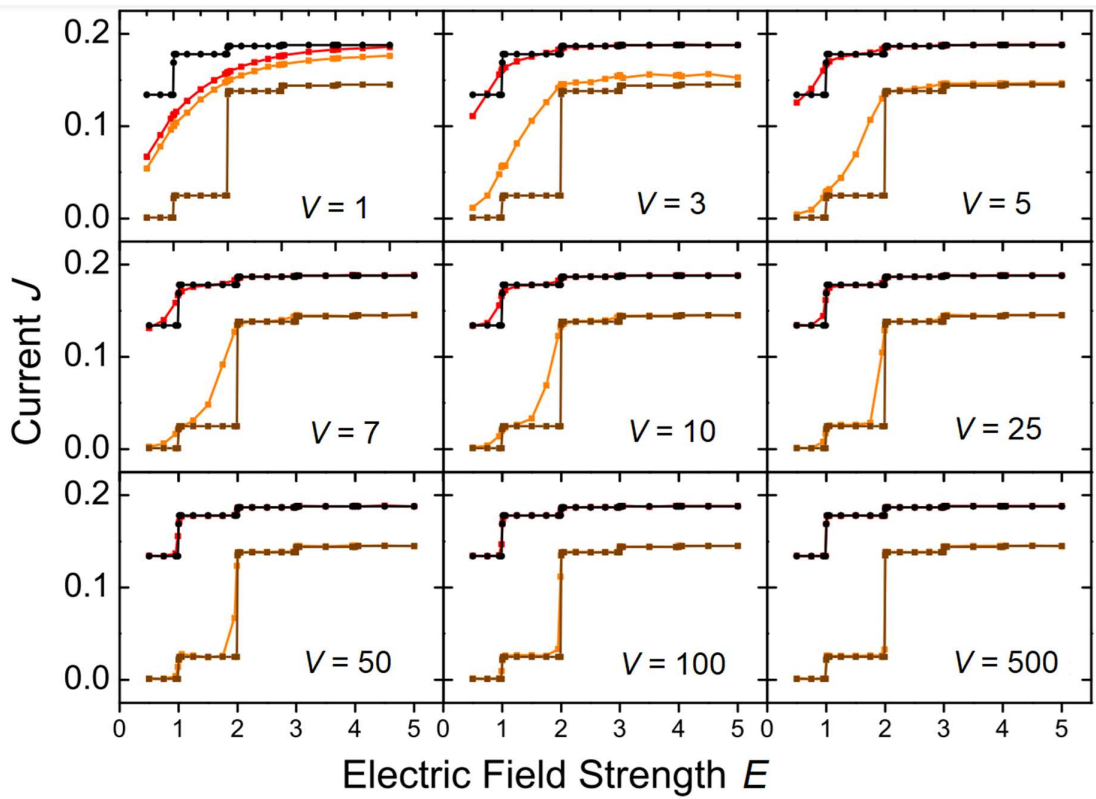

**Figure 3**. The plot of current versus electric field strength, obtained from the MC simulation. Black and brown curves indicate $V = \infty$ for repulsive and attractive interactions, respectively. Centered curves (red and orange on online) represent the results at each V value for repulsive and attractive interactions, respectively.

Figure 4 shows the results of the second assumption. In this case, $U$ of the second-neighbor was taken as 0.5; therefore, the interaction energy between particles is increased. Due to the additional interaction derived from the second neighbors, additional steps appeared at $E = 0.5$, 1.5, and 2.5. The curves do not overlap with decreasing V values.

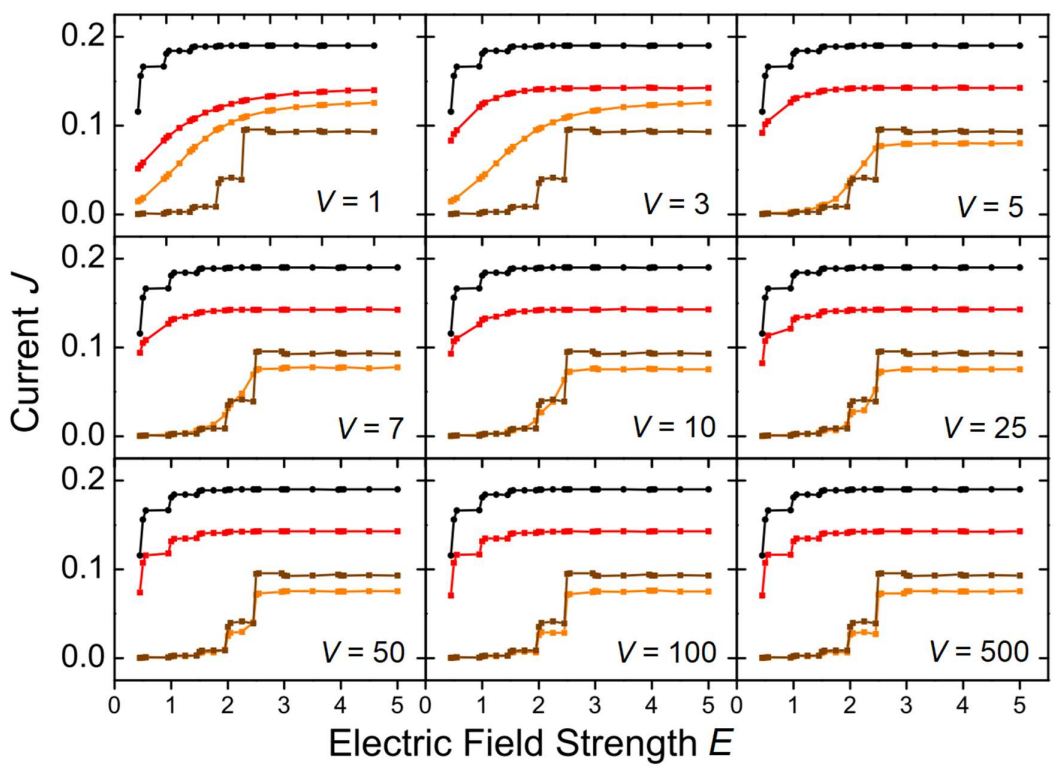

**Figure 4**. The plot of current versus electric field strength, obtained from the MC simulation for second-neighbor interaction. Top and bottom curves (black and brown on online) curves indicate $V = \infty$ for repulsive and attractive interactions, respectively. Centered curves (red and orange on online) curves represent the results at each T value for repulsive and attractive interactions, respectively.

**Concluding remarks**

By increasing the electric field strength $E$, a smooth flow, i.e., the increase in current $J$ is obtained, as expected. On the other hand, when interactions between particles are increased (which means $V = \infty$) a phase transition and stepwise increase in the current is observed. The stepwise increase is also related to the number of interactions, for instance when the second nearest neighbor interactions are included additional transitions at half numbers were obtained. This shows that stronger interaction between particles will be leading to the aggregation of the molecules and the fading of the individual movements.

This study reports a modeling study of the movement of molecules in a capillary column under electroosmotic flow for the first time. Using the approach demonstrated in this study, further studies may be useful for explaining or predicting the electrophoretic behaviors of mixtures of different type of molecules, including the entanglement of segmented macromolecules, flows in coated columns, and/or gel media. By adjusting the electric field strength, desired degree of movement can be obtained and a phase transition behavior in the capillary can be obtained.

**Acknowledgments**

A. Nihat Berker gratefully acknowledges support by the Academy of Sciences of Turkey (TÜBA).

**Conflict of interest**

The authors have declared no conflict of interest.